\documentclass{article}
\usepackage{spconf,amsmath,graphicx,hyperref}
\usepackage{cite}
\usepackage{amsmath,amssymb,amsfonts}
\usepackage{booktabs}
\usepackage{hyperref}
\usepackage{multirow}
\usepackage{algorithmic}
\usepackage{graphicx}
\usepackage{textcomp}
\usepackage{xcolor}
\usepackage{url}
\usepackage{colortbl} 
\usepackage{booktabs}
\usepackage{threeparttable}
\usepackage{makecell}
\usepackage{svg}
\usepackage[x11names]{xcolor}
\usepackage{float}
\usepackage{acronym}
\definecolor{mygreen}{RGB}{34,139,34}   
\usepackage{arydshln} 

\acrodef{ssl}[SSL]{Self-Supervised Learning}
\acrodef{mir}[MIR]{Music Information Retrieval}
\acrodef{qa}[QA]{Question-Answering}

\ninept

\title{TinyMU: A Compact Audio-Language Model for Music Understanding}
\name{Xiquan Li$^{1,2}$, Aurian Quelennec$^{1}$, Slim Essid$^{1}$ 
        \thanks{S. Essid is now with NVIDIA. The work was performed when he was with Télécom Paris. This work was supported by the Audible project, funded by BPI. Codes and data are available at \url{https://github.com/xiquan-li/TinyMU}. } 
}
\address{$^{1}$\textit{LTCI}, \textit{Télécom Paris}, \textit{Institut Polytechnique de Paris} \\
          $^{2}$\textit{Shanghai Jiao Tong University}
          }
\begin{document}
\maketitle
\begin{abstract}
Music understanding and reasoning are central challenges in the Music Information Research field, with applications ranging from retrieval and recommendation to music agents and virtual assistants. 
Recent Large Audio-Language Models (LALMs) have shown remarkable progress in answering music-related questions by following user instructions. 
However, their massive scale, often billions of parameters, results in expensive training, slow inference, and limited deployability on edge devices.

In this work, we present TinyMU, a lightweight (229M) Music-Language Model (MLM) that achieves performance comparable to much larger LALMs while remaining efficient and compact. 
To train TinyMU, we introduce MusicSkills-3.5M, a carefully curated, music-grounded question-answering dataset with 3.5M samples. Spanning multiple-choice, binary, and open-ended formats, this dataset provides fine-grained supervision across diverse musical concepts. 
For its architecture, TinyMU leverages MATPAC++, the SOTA self-supervised audio encoder for fine-grained feature extraction. 
Paired with a lightweight linear projector, it efficiently aligns audio embeddings with the language model. 
Through extensive evaluation, we show that TinyMU performs strongly in both basic music understanding and complex reasoning. 
Notably, on the MuChoMusic benchmark, it achieves 82\% of SOTA LALM's performance despite being 35× smaller, highlighting the potential of small MLMs under constrained computational budgets.

\end{abstract}

\section{Introduction}

Recent years have witnessed rapid progress in Large Audio Language Models (LALMs) \cite{chu2023qwen, chu2024qwen2, gong2023listen, tang2023salmonn, ghosh2025audio}, which integrate robust audio encoders with Large Language Models (LLMs). Trained on large-scale audio–text datasets, LALMs demonstrate impressive understanding and reasoning abilities, moving beyond traditional tasks such as tagging or classification to describing and reasoning about complex audio scenes.
Within this trend, Music Language Models (MLMs) \cite{deng2023musilingo, liu2024music, zhao2024openmu, gardner2023llark} have emerged as a specialized branch focusing on music understanding. They can handle classic music information retrieval (MIR) tasks, as well as more challenging ones, such as music captioning and open-domain music question answering.

In pursuit of stronger MLMs, recent studies have explored three complementary directions: architectural innovations \cite{deng2023musilingo, liu2024music}, scaling up data and model parameters \cite{zhao2024openmu, chu2023qwen, chu2024qwen2}, and designing new training objectives \cite{li2025reinforcement}. These efforts have led to remarkable gains: the average accuracy on the MuChoMusic benchmark \cite{weck2024muchomusic} has surged from 21\% \cite{deng2023musilingo} to 71\% \cite{dinkel2025midashenglm} in just a few years.

However, these improvements have come at the cost of model size. State-of-the-art Music Language Models such as MiDashengLM, Qwen2-Audio and Audio-Flamingo 3 \cite{dinkel2025midashenglm, chu2024qwen2, goel2025audioflamingo3} exceed 8 billion parameters, making them expensive to train, slow to run, and impractical for deployment on edge devices. 
Despite the clear importance of efficiency, low latency, and accessibility, relatively little research has focused on small-scale Music Language Models, which could bring music reasoning capabilities to real-time, on-device, and resource-constrained environments.

\begin{figure}[t]
\centering
\includegraphics[width=\linewidth]{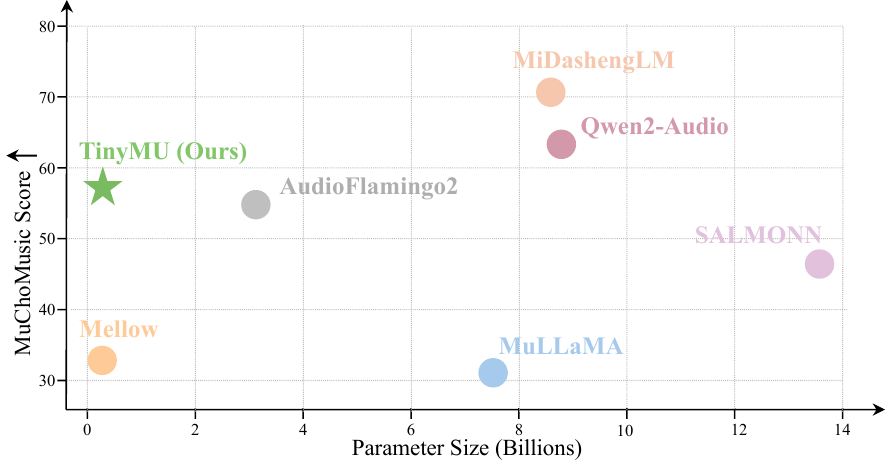}
\vspace{-0.5cm}
\caption{TinyMU achieves 82\% of the performance of SOTA LALMs in MuChoMusic benchmark, while using 35 times fewer parameters.}
\label{fig:tinymu}
\vspace{-0.5cm}
\end{figure}

In this paper, we present TinyMU, a compact MLM that achieves strong music understanding and reasoning capabilities while being highly efficient.
With only 229M parameters, TinyMU performs on par with many recent LALMs. 
To support training, we first introduce MusicSkills-3.5M, a comprehensive music-grounded question-answering dataset with diverse formats.
Specifically, we create multiple-choice, binary, and open-ended \ac{qa} samples that cover a broad spectrum of music knowledge and reasoning, including theory, performance, structure, etc. 
Such data diversity proves instrumental in boosting the model’s perceptual and reasoning capabilities. 
For its architecture, TinyMU leverages MATPAC++ \cite{quelennec2025matpac++}, a state-of-the-art self-supervised pre-trained audio encoder to capture fine-grained musical details. 
A lightweight linear projector then maps the audio embeddings into the language space, enabling efficient alignment with the Small Language Model (SLM). 

Extensive experiments show that TinyMU performs competitively on fundamental music understanding tasks such as captioning, instrument recognition, and genre classification, while also demonstrating strong results on challenging reasoning benchmarks. 
In addition, a comprehensive ablation study sheds light on the most effective strategies for training compact MLMs, highlighting the role of data diversity and architectural choices.

To summarize, our main contributions are as follows:
(i) We propose TinyMU, a compact Music Language Model (229M parameters) that delivers strong performance on both music understanding and reasoning tasks, rivaling with much larger LALMs.
(ii) We introduce MusicSkills-3.5M, a large and diverse music-grounded QA dataset spanning multiple-choice, binary, and open-ended formats.
(iii) We conduct extensive comparisons and ablation studies, establishing best practices for designing efficient small-scale Music Language Models.


\section{The MusicSkills-3.5M Dataset}
\label{sec:musicskill}
As a foundation for TinyMU, we construct MusicSkills-3.5M, a diverse music question-answering dataset designed to enhance both the understanding and reasoning capabilities of efficient Music Language Models.
Following existing works \cite{gong2023listen, tang2023salmonn, gardner2023llark}, we assemble a collection of samples consisting of \texttt{(Music, Question, Answer)} pairs.

Unlike existing datasets that primarily rely on open-ended formats, we hypothesize that incorporating a diversity of question types should better strengthen the alignment between music content and textual queries. Open-ended \texttt{Questions}, such as “What is happening in the music?”, with relatively unconstrained \texttt{Answers}, provide rich but diffuse information that can be difficult for the model to interpret.
Therefore, we additionally curated multiple-choice questions, which provide precise queries with explicitly defined correct and incorrect answers, thereby offering highly informative signals for learning fine-grained distinctions. We further incorporated binary (yes/no) questions, which are strongly directive and guide the model to associate specific musical cues with factual properties of the audio. Taken together, this mixture of formats enables the model to ground its representations more effectively in the underlying music signals.
This section details our data creation process.

\begin{table}[t]

\centering
 \resizebox{\linewidth}{!}{
\begin{tabular}{lclllll}
\toprule
\multirow{2.5}{*}{\textbf{Audio Sources}} & \multirow{2.5}{*}{\textbf{Audios}} & \multicolumn{4}{c}{\textbf{Tasks}} & \multirow{2.5}{*}{\textbf{Total}} \\ \cmidrule{3-6}
& &  \textbf{Captioning} & \textbf{QA} & \textbf{MCQ} & \textbf{Binary} \\
\midrule
MusicCaps & 2.2k & 13k & 42k &30k  & 13k  & 98k \\
MagnaTagATune   & 17k & 62k &62k & 162k  & 62k   & 348k \\
FMA & 172k & 172k & 688k & 775k & 258k & 1.9M \\
AudioSet  & 317k  & 317k & 317k  & 316k  & 317k  & 1.2M  \\
\midrule
Total & 508k & 564k & 1.2M & 1.2M & 650k & 3.5M \\
\bottomrule
\end{tabular}
}
\vspace{-0.25cm}
\caption{Statistics of MusicSkills-3.5M}
\vspace{-0.5cm}
\label{tab:musicskill}
\end{table}

\subsection{Music Sources}
We primarily select four datasets as music sources: MusicCaps \cite{agostinelli2023musiclm}, MagnaTagATune (MTT) \cite{law2009evaluation}, FMA \cite{defferrard2016fma}, and AudioSet \cite{gemmeke2017audio}.
MusicCaps, MagnaTagATune, and FMA are widely adopted in MIR research, as they provide professional music recordings annotated with rich labels and descriptive captions \cite{doh2023lp, lanzendorfer2025coarse}.
In contrast, AudioSet has rarely been exploited for music understanding. Nevertheless, nearly half of its clips are associated with music-related tags, and many consist of real-life performances (e.g., people playing instruments in natural settings). Coupled with its comprehensive ontology, these characteristics make AudioSet particularly well-suited for constructing diverse music question–answering data.

\subsection{Data Construction}
We use both a rule-based method and an LLM-assisted method to synthesize our music-related QA data, which we detail below. 

\noindent \textbf{Rule-based QA Data Creation.}
For AudioSet, we leverage its hierarchical ontology \cite{gemmeke2017audio} and define rules to create QA pairs in open-ended, binary, and multiple-choice formats. 
Specifically, we begin by selecting audio clips that contain at least one music-related leaf node label from the ontology. Audio clips that only have parent-level labels were excluded, since these labels (e.g. music, music instrument) can only indicate the general presence of music and do not provide enough information for QA construction. 
This filtering process resulted in about 300k clips, each tagged with at least one fine-grained label (e.g., Acoustic Guitar, Reggae, etc.). 
Then, we generated diverse QA instances by using each leaf label along with its parent category. 
For instance, for music that has the label \textit{Acoustic Guitar}, we can utilize its parent label \textit{instrument} and create questions such as: \textit{What is the \{instrument\} in this music?} (Open-ended QA), \textit{Is \{Guitar\} present in the music?} (Binary QA), and \textit{Select the \{instrument\} in this music: A. Guitar, B. Violin, C. Ukulele, D. Cello.} (Multiple-choice QA). 
For each parent label, we prompt ChatGPT to generate several question templates and randomly select one during data creation. 
For distractors of multiple-choice QAs, we sample labels based on their occurrence frequency in the dataset. Consequently, common labels (e.g., Drums, Guitar) are more likely to appear as distractors. We hypothesize that this sampling strategy could increase task difficulty and compel the model to develop a finer understanding of musical details.
Through the rule-based method, we obtained around 1M QA pairs, centered on genre, mood, instrumentation, and musical role.

\begin{figure}[t]
    \centering
    \vspace{-0.3cm}
    \includegraphics[width=\linewidth]{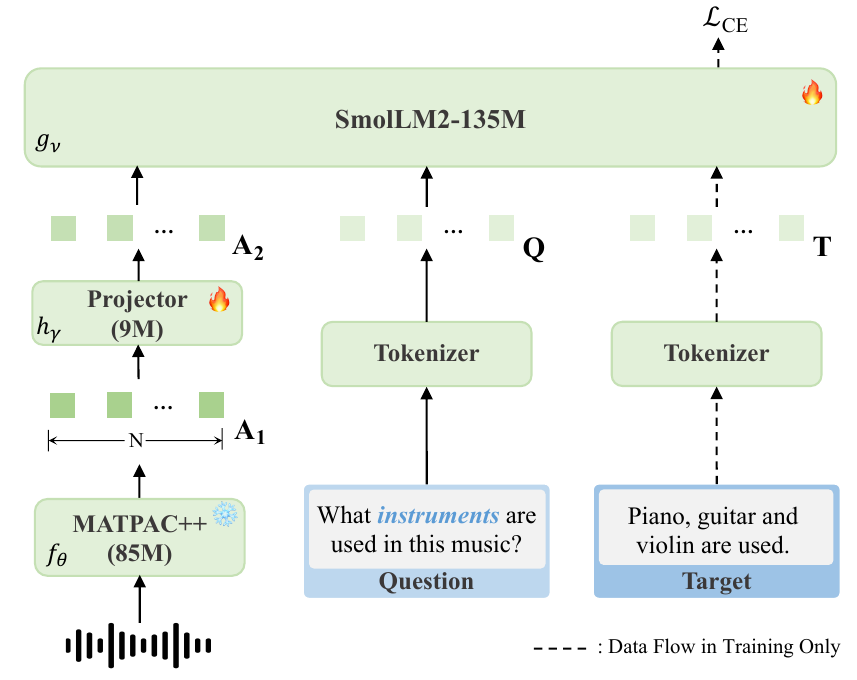}
    \vspace{-0.5cm}
    \caption{The overall architecture of TinyMU }
    \label{fig:arch_tinymu}
    \vspace{-0.5cm}
\end{figure}

\begin{table*}[t]
\centering
\resizebox{1\linewidth}{!}{
\renewcommand{\arraystretch}{1.1} 
\begin{tabular}{lllllllll}
\toprule
\multirow{2.5}{*}{\textbf{Method}} &  \multirow{2.5}{*}{\textbf{Size}} & \multicolumn{2}{c}{\textbf{Basic MIR Tasks  (\%)}} & \multicolumn{2}{c}{\textbf{MusicCaps (\%)}} & \multicolumn{3}{c}{\textbf{MuChoMusic (\%)}} \\
\cmidrule(lr){3-4} \cmidrule(lr){5-6} \cmidrule(lr){7-9}
 &  &  GTZAN  & Medley-Solos-DB & METEOR & BERT-Score & Knowledge & Reasoning & All    \\
\midrule
MusiLingo & 7.1B &  57.7 & 30.5 & \underline{21.7} & 86.8 & 33.6 & 28.2 & 31.5 \\
MU-LLaMA & 7.7B & 37.3 & 38.6 & 12.3 & 86.8 & 32.3 & 33.5 & 32.7 \\
Audio-Flamingo 2 
& 4.4B & 69.1 & 85.6  & 13.3 & 86.1 & - & - & 56.5 \\
MiDashengLM
& 8.3B & 72.7 & \underline{85.8} & 14.8 & 87.3 & - & - & \textbf{71.4} \\
Audio-Flamingo 3 & 8.3B & \textbf{83.2} & 83.4 & 11.8 & \underline{87.8} & - & - & 47.4  \\
Qwen2-Audio-Instruct 
& \textbf{8.4B} & \underline{77.2} & 80.3 & \textbf{23.4} & \textbf{88.2} & \textbf{69.4} & \textbf{65.5} & \underline{67.8} \\
\hdashline
Mellow & 167M & 16.5 & 49.6 & 15.0 & 85.8 & 30.8 & 32.0 & 30.3 \\
\rowcolor{blue!10}
TinyMU (Ours) & 229M\textcolor{mygreen}{\scriptsize (2.7\%)} 
& 65.7\textcolor{mygreen}{\scriptsize (78.9\%)} 
& \textbf{95.1}\textcolor{mygreen}{\scriptsize (100.0\%)} 
& 16.9\textcolor{mygreen}{\scriptsize (72.2\%)} 
& 87.3\textcolor{mygreen}{\scriptsize (99.0\%)} 
& \underline{58.3}\textcolor{mygreen}{\scriptsize (84.0\%)} 
& \underline{59.6}\textcolor{mygreen}{\scriptsize (91.0\%)} 
& 58.6\textcolor{mygreen}{\scriptsize (82.1\%)} \\
\bottomrule 
\end{tabular}
}
\vspace{-0.2cm}
\caption{Overall performance comparison between TinyMU and state-of-the-art audio language models. 
The best results are highlighted in \textbf{bold}, and the second-best results are \underline{underlined}. Alongside TinyMU’s performance, we also report its \textcolor{mygreen}{percentage} relative to SOTA results.
}
\label{tab:overall_results}
\vspace{-3mm}
\end{table*}

\noindent \textbf{LLM-Assisted QA Data Creation.}
For MusicCaps, Magna-Tag-A-Tune (MTT) and FMA, we use LLM-based generation methods to synthesise QA pairs. 
Following prior works \cite{gong2023listen, tang2023salmonn, zhao2024openmu}, we leverage a textual language model with metadata and captions to generate music-related QA. To improve diversity and depth, we design prompts that elicit questions spanning a wide range of musical knowledge and reasoning aspects.
Specifically, we first define key dimensions of music understanding, including instrumentation, melody, tempo, genre, mood, function, etc. We then construct illustrative examples in open-ended, binary, and MCQ formats for each dimension. 
Leveraging the strong in-context learning capabilities of the LLM, we use prompts, examples, and given metadata of the music clip to produce reasoning-oriented QA pairs. 
In total, the LLM-based method yields approximately 2M QA instances. 
While LLM-based methods can generate more diverse QA requiring higher levels of reasoning, rule-based methods tend to produce more fundamental, perception-oriented questions. By combining the two, we aim to enhance both the perceptual and reasoning abilities of TinyMU.

To further enrich our dataset, we also incorporate existing music captioning datasets: MusicCaps \cite{agostinelli2023musiclm}, LP-MusicCaps \cite{doh2023lp}, AudioSetCaps \cite{bai2024audiosetcaps}, along with two open-ended music QA datasets: MusicInstruct \cite{deng2023musilingo} and OpenMU-MTT \cite{zhao2024openmu}, into MusicSkills. 
Table \ref{tab:musicskill} summarizes the overall statistics of the constructed dataset, including the number of samples for different tasks and their distribution across audio sources. In total, we collect 3.5M question–answer pairs, covering diverse aspects of music and spanning three major QA formats.


\section{TinyMU}
\noindent \textbf{Overview.}
As illustrated in Figure \ref{fig:arch_tinymu}, TinyMU is composed of three core components: 
An audio encoder $f_\theta$ to extract fine-grained music features, a projector $h_\gamma$ to align different modalities, and a small language model $g_\nu$ to generate text conditioned on music and user instructions. 

\noindent \textbf{Encoder.}
TinyMU leverages the audio encoder MATPAC++ \cite{quelennec2025matpac++} to encode input music clips. 
MATPAC++ is the state-of-the-art self-supervised pre-trained audio encoder that leverages Multiple-Choice Learning (MCL) with masked latent prediction. It demonstrates strong capabilities in music representation learning, outperforming numerous audio encoders such as MERT \cite{li2023mert} and BEATs \cite{chen2022beats}. 
Given an audio clip, MATPAC++ first converts the raw waveform into a log-scale Mel spectrogram. This spectrogram is partitioned into non-overlapping patches and processed by the transformer backbone, yielding fine-grained embeddings $\mathbf{A}_1 \in \mathbb{R}^{N \times d_1}$, where $N$ and $d_1$ denote the embedding length and dimension, respectively.

\noindent \textbf{Projector.}
We then feed $\mathbf{A}_1' \in \mathbb{R}^{N \times  d_1}$ into the projector $h_\mu: \mathbb{R}^{d_1} \rightarrow \mathbb{R}^{d_2}$, resulting in $\mathbf{A}_2 \in \mathbb{R}^{N\times d_2}$, where $d_2$ is the embedding dimension of the language model. We use a simple projector $h_\mu$ comprising two linear layers. 
Early experiments indicate that a lightweight projector is sufficient for aligning audio and language representations in TinyMU, and that increasing its complexity does not lead to measurable gains.

\noindent \textbf{Language Model.}
We adopt SmolLM2 \cite{allal2025smollm2}, one of the strongest small language models, as our language model $g_\nu$. 
During training, we feed the constructed \texttt{question}–\texttt{answer} pairs into the tokenizer of $g_\nu$, obtaining $\textbf{Q}, \textbf{T}$. 
We then train TinyMU to minimize the cross-entropy loss between the prediction and the target conditioned on the encoded question $\mathbf{Q}$ and audio embedding $\mathbf{A}_2$: 
$$
\mathcal{L}_\text{CE} = -\frac{1}{|\mathbf{T}|}\sum_{i=1}^{|\mathbf{T}|}\log p (\mathbf{T}_i| \mathbf{A}_2, \mathbf{Q}) \; ;
$$
where $|\mathbf{T}|$ denotes the length of the target text, and $\mathbf{T}_i$ is its $i$-th token. During training, we freeze the audio encoder and only train the projector along with the language model.

\begin{table*}[t]
\centering
\resizebox{1\linewidth}{!}{
\renewcommand{\arraystretch}{1.1} 
\begin{tabular}{lllccccccc}
\toprule
\multirow{2.5}{*}{\textbf{Method}} &  \multirow{2.5}{*}{\textbf{Encoder}} & \multirow{2.5}{*}{\textbf{LLM Tuning}} & \multicolumn{2}{c}{\textbf{Basic MIR Tasks  (\%)}} & \multicolumn{2}{c}{\textbf{MusicCaps (\%)}} & \multicolumn{3}{c}{\textbf{MuChoMusic (\%)}} \\
\cmidrule(lr){4-5} \cmidrule(lr){6-7} \cmidrule(lr){8-10}
 &  &  & GTZAN  & Medley-Solos-DB & METEOR & BERT-Score & Knowledge & Reasoning & All    \\
\midrule
\rowcolor{blue!10}
TinyMU (Ours) & MATPAC++ & Full-Tuning & \textbf{65.7} & \textbf{95.1} & 16.9 & 87.3 & \textbf{58.3} & \textbf{59.6} & \textbf{58.6} \\
\quad w. HTSAT & HTSAT & Full-Tuning & 60.6 & 64.7 & 17.2 & 87.3 & 54.4 & 56.9 & 55.2 \\
\quad w. Frozen LLM & MATPAC++ & Frozen & 43.1 & 25.4 & \textbf{17.5} & 86.3 & 23.4 & 25.7 & 24.2 \\
\quad w. LoRA ($r, \alpha$) & MATPAC++ & LoRA (8, 32) & 55.0 & 87.3 & 16.5 & 87.3  & 36.7 & 45.9 & 39.9 \\
\quad w. LoRA ($r, \alpha$) & MATPAC++ & LoRA (32, 128) & 56.1 & 89.1 & 17.2 & \textbf{87.4} & 44.2 & 50.9 & 46.3 \\
\bottomrule 
\end{tabular}
}
\vspace{-0.3cm}
\caption{Ablation study of different audio encoders and training strategies. }

\label{tab:ablation_training}
\vspace{-3mm}
\end{table*}

\section{Experiments}
\subsection{Evaluation Setup}
We conducted a holistic evaluation to assess TinyMU's music understanding and reasoning capabilities. The evaluation involves three distinct levels of tasks: 

\noindent\textbf{Basic Music Information Retrieval (MIR) tasks. }We use GTZAN \cite{george2001automatic} and Medley-Solos-DB \cite{lostanlen2019medley} to evaluate model's fundamental music understanding. GTZAN is a genre classification dataset consisting of 1,000 clips across 10 genres. Medley-Solos-DB is an instrument recognition benchmark with 12,236 solo clips, each corresponding to one of eight instrument categories. Since neither dataset overlaps with TinyMU’s training data, they serve as reliable benchmarks for zero-shot perceptual evaluation.
Following prior work \cite{ghosh2025audio, gong2023listen}, we compute the similarity between TinyMU’s outputs and candidate labels using the text encoder of CLAP \cite{elizalde2023clap}, and select the highest-scoring label as model output. 
We then report classification accuracy for these two benchmarks. 

\noindent \textbf{Music Captioning. }We evaluate TinyMU on the test split of MusicCaps to assess its ability to generate descriptive captions for music.
MusicCaps is the standard benchmark for music captioning, containing 2.8k music clips in its test set. 
We report the commonly adopted metrics: METEOR \cite{banerjee2005meteor}, and BERTScore \cite{zhang2019BERTScore}. 

\noindent \textbf{Complex Music Reasoning. }We leverage MuChoMusic \cite{weck2024muchomusic}, a widely used and comprehensive benchmark designed for music reasoning evaluation. MuChoMusic comprises 1,187 multiple-choice questions spanning 23 distinct musical dimensions, including melody, harmony, rhythm, texture, genre, etc. This makes it a robust benchmark to assess both the understanding and the complex reasoning capabilities of Music-Language Models. 
We report the accuracy of models on the benchmark.


\subsection{Main Results}
We compare TinyMU against state-of-the-art audio language models, including Mellow \cite{deshmukh2025mellow}, MU-LLaMA \cite{liu2024music}, MusiLingo \cite{deng2023musilingo}, Audio-Flamingo series \cite{ghosh2025audio, goel2025audioflamingo3}, MiDashengLM \cite{dinkel2025midashenglm}, and Qwen2-Audio-Instruct \cite{chu2024qwen2}. 

Among them, Mellow is another small audio-language model with only 167M parameters. 
MU-LLaMA and MusiLingo are Large Music Language Models, with training datasets primarily composed of music-related questions.
Audio-Flamingo series, MiDashengLM, and Qwen2-Audio-Instruct are large-scale LALMs that have been trained on more diverse audio data corpora.
During evaluation, we compute metrics that are not reported in the original papers using the officially released checkpoints and assess the outputs of the full multi-modal models.

As shown in Table \ref{tab:overall_results}, TinyMU achieves 65.7\% on GTZAN and 95.1\% on Medley-Solos-DB, demonstrating strong music perception and outperforming all baselines on instrument recognition.
On MusicCaps, it records 16.9 METEOR and 87.3 BERT-Score, corresponding to 72.2\% and 99.0\% of the best systems, indicating its ability to generate semantically faithful music descriptions.
On MuChoMusic, TinyMU achieves 58.6 overall, reaching 82\% of the performance of MiDashengLM while being 35$\times$ smaller. 
Notably, it significantly surpasses other small audio–language models and achieves performance comparable to larger models, including MusiLingo, MU-LLaMA, and Audio-Flamingo, underscoring its strong capacity for music understanding and reasoning.


\subsection{Ablation Studies}

To explore best practices for building compact music language models, we conducted a comprehensive ablation study covering architecture design, training strategies, and data construction.

\noindent \textbf{Audio Encoders. }We first study the impact of different audio encoders by replacing MATPAC++ with HTS-AT \cite{chen2022hts}, which is another powerful audio encoder widely used by many other audio-language models \cite{elizalde2023clap, deshmukh2025mellow}. 
As illustrated in Table \ref{tab:ablation_training}, replacing MATPAC++ with HTSAT leads to notable performance degradation. Notably, the model performs much worse on Medley-Solos-DB (64.7\% vs. 95.1\%) and achieves lower reasoning accuracy on MuChoMusic (56.9\% vs. 59.6\%). 
We attribute this gap to the self-supervised training strategy of MATPAC++, which yields audio representations that are less task-oriented, compared to HTSAT, which is trained for AudioSet classification.

\noindent \textbf{Training Strategies. }
We then explore the best fine-tuning methods for the language model, including complete freezing, LoRA \cite{hu2022lora} with different hyperparameters, and full fine-tuning.

As shown in Table \ref{tab:ablation_training}, completely freezing the language model and only training the linear projector yields relatively strong results in music captioning. Specifically, the model achieves a METEOR of 17.5, demonstrating a better lexical overlap with ground-truth captions. 
However, the model suffers from severe degradation in music question answering, as its accuracy drops to 43.1\% on GTZAN, 25.4\% on Medley-Solos-DB, and 25.7\% on MuChoMusic. 
These results suggest that employing a simple linear mapper is effective in equipping a small language model with the capacity to describe music.
However, for tasks requiring deeper music reasoning and knowledge, leaving the SLM unadapted leads to sub-optimal performance, which may stem from the limited musical knowledge and insufficient depth of understanding in small language models.


When applying LoRA to fine-tune the language model, TinyMU shows clear gains in both basic music understanding and complex reasoning.
For instance, using LoRA with $(r=8, \alpha=32)$ yields an accuracy of 55.0 on GTZAN and 39.9 on MuChoMusic, representing a substantial improvement over the frozen baseline.
Furthermore, increasing the number of trainable parameters enhances the model’s comprehension ability. For example, using $r=32$ and $\alpha=128$ for LoRA 
raises the accuracy to 89.1 on Medley and 46.3 on MuCho, substantially outperforming the smaller configuration $(r=8, \alpha=32)$. 
Finally, full fine-tuning of the language model yields the strongest QA performance, confirming that greater adaptation capacity leads to superior reasoning ability.

\noindent \textbf{QA Types. }
We then study the contribution of different data formats by selectively removing individual QA types from the full MusicSkills-3.5M dataset. 
As shown in Table~\ref{tab:ablation_data}, removing open-ended QA leads to substantial drops in both perception and reasoning tasks, particularly on GTZAN (–14.5) and Medley-Solos-DB (–15.5). This highlights the critical role of open-ended QA in helping the model recognize and generalize across diverse musical expressions, which is essential for perception tasks.
Secondly, excluding binary QA results in consistent performance drops across all three benchmarks (–2.0 on GTZAN, –0.3 on Medley, and –2.2 on MuChoMusic), highlighting its role in linking audio cues to factual properties and enhancing reasoning.
Thirdly, removing the MCQ component leads to a dramatic performance drop on MuChoMusic (–34.1 points), which is expected given the MCQ-based nature of the benchmark. Notably, MCQ-style supervision also improves basic MIR and captioning tasks, as it encourages fine-grained discrimination between correct answers and distractors. 
Finally, combining all three types of QA yields the strongest performance, underscoring the necessity of diverse supervision strategies in building effective small music language models.

\begin{table}[t]
\centering
\resizebox{1\linewidth}{!}{
\renewcommand{\arraystretch}{1.1} 
\begin{tabular}{lccc}
\toprule
\textbf{Training Data} & \textbf{GTZAN} & \textbf{Medley-Solos-DB} & \textbf{MuChoMusic} \\
\midrule
\rowcolor{blue!10}
MusicSkills-3.5M & \textbf{65.7} & \textbf{95.1}  & \textbf{58.6} \\
\quad w/o Open-ended QA & 51.2 & 79.6 & 54.9 \\
\quad w/o Binary QA & 63.7 & 94.8 & 56.4 \\
\quad w/o MCQ  & 65.0 & 93.4 & 24.5 \\

\bottomrule 
\end{tabular}
}
\vspace{-0.3cm}
\caption{Ablation study of different QA formats. }

\label{tab:ablation_data}
\vspace{-3mm}
\end{table}










\section{Conclusion}
This paper proposes TinyMU, a compact Music-Language Model with strong understanding and reasoning abilities. To train TinyMU, we introduce MusicSkills-3.5M, a comprehensive music-grounded dataset spanning diverse formats and musical concepts. 
To boost perception, we leverage MATPAC++, the SOTA self-supervised audio encoder as TinyMU's feature extractor. 
Experimental results show that TinyMU performs competitively with larger audio–language models, despite being more than an order of magnitude smaller. 
Comprehensive ablation studies further reveal effective strategies for training small MLMs and constructing high-quality music QA datasets, paving the way toward more accessible and powerful systems for music understanding.


\bibliographystyle{IEEEtran} 
\bibliography{refs}

@article{gong2023listen,
  title={Listen, think, and understand},
  author={Gong, Yuan and Luo, Hongyin and Liu, Alexander H and Karlinsky, Leonid and Glass, James},
  journal={Proc. ICLR},
  year={2024}
}

@article{tang2023salmonn,
  title={{SALMONN}: Towards generic hearing abilities for large language models},
  author={Tang, Changli and Yu, Wenyi and Sun, Guangzhi and Chen, Xianzhao and Tan, Tian and Li, Wei and Lu, Lu and Ma, Zejun and Zhang, Chao},
  journal={Proc. ICLR},
  year={2024}
}

@article{chu2023qwen,
  title={Qwen-audio: Advancing universal audio understanding via unified large-scale audio-language models},
  author={Chu, Yunfei and Xu, Jin and Zhou, Xiaohuan and Yang, Qian and Zhang, Shiliang and Yan, Zhijie and Zhou, Chang and Zhou, Jingren},
  journal={arXiv preprint arXiv:2311.07919},
  year={2023}
}

@article{chu2024qwen2,
  title={Qwen2-audio technical report},
  author={Chu, Yunfei and Xu, Jin and Yang, Qian and Wei, Haojie and Wei, Xipin and Guo, Zhifang and Leng, Yichong and Lv, Yuanjun and He, Jinzheng and Lin, Junyang and others},
  journal={arXiv preprint arXiv:2407.10759},
  year={2024}
}

@article{zhao2024openmu,
  title={{OpenMU}: Your Swiss Army Knife for Music Understanding},
  author={Zhao, Mengjie and Zhong, Zhi and Mao, Zhuoyuan and Yang, Shiqi and Liao, Wei-Hsiang and Takahashi, Shusuke and Wakaki, Hiromi and Mitsufuji, Yuki},
  journal={arXiv preprint arXiv:2410.15573},
  year={2024}
}

@inproceedings{lanzendorfer2025coarse,
  title={Coarse-to-Fine Text-to-Music Latent Diffusion},
  author={Lanzend{\"o}rfer, Luca A and Lu, Tongyu and Perraudin, Nathana{\"e}l and Herremans, Dorien and Wattenhofer, Roger},
  booktitle={Proc. ICASSP},
  year={2025},
}

@article{weck2024muchomusic,
  title={{MuChoMusic}: Evaluating music understanding in multimodal audio-language models},
  author={Weck, Benno and Manco, Ilaria and Benetos, Emmanouil and Quinton, Elio and Fazekas, George and Bogdanov, Dmitry},
  journal={Proc. ISMIR},
  year={2024}
}

@article{allal2025smollm2,
  title={{SmolLM2}: When Smol Goes Big--Data-Centric Training of a Small Language Model},
  author={Allal, Loubna Ben and Lozhkov, Anton and Bakouch, Elie and Bl{\'a}zquez, Gabriel Mart{\'\i}n and Penedo, Guilherme and Tunstall, Lewis and Marafioti, Andr{\'e}s and Kydl{\'\i}{\v{c}}ek, Hynek and Lajar{\'\i}n, Agust{\'\i}n Piqueres and Srivastav, Vaibhav and others},
  journal={arXiv preprint arXiv:2502.02737},
  year={2025}
}

@inproceedings{chen2022hts,
  title={{HTS-AT}: A hierarchical token-semantic audio transformer for sound classification and detection},
  author={Chen, Ke and Du, Xingjian and Zhu, Bilei and Ma, Zejun and Berg-Kirkpatrick, Taylor and Dubnov, Shlomo},
  booktitle={Proc. ICASSP},
  year={2022},
}

@article{deng2023musilingo,
  title={Musilingo: Bridging music and text with pre-trained language models for music captioning and query response},
  author={Deng, Zihao and Ma, Yinghao and Liu, Yudong and Guo, Rongchen and Zhang, Ge and Chen, Wenhu and Huang, Wenhao and Benetos, Emmanouil},
  journal={Proc. NAACL},
  year={2024}
}

@inproceedings{liu2024music,
  title={Music understanding {LLaMA}: Advancing text-to-music generation with question answering and captioning},
  author={Liu, Shansong and Hussain, Atin Sakkeer and Sun, Chenshuo and Shan, Ying},
  booktitle={Proc. ICASSP},
  year={2024},
}

@inproceedings{goel2025audioflamingo3,
      title={Audio Flamingo 3: Advancing Audio Intelligence with Fully Open Large Audio Language Models}, 
      author={Arushi Goel and Sreyan Ghosh and Jaehyeon Kim and Sonal Kumar and Zhifeng Kong and Sang-gil Lee and Chao-Han Huck Yang and Ramani Duraiswami and Dinesh Manocha and Rafael Valle and Bryan Catanzaro},
      year={2025},
      booktitle={arXiv},
}

@article{ghosh2025audio,
  title={Audio flamingo 2: An audio-language model with long-audio understanding and expert reasoning abilities},
  author={Ghosh, Sreyan and Kong, Zhifeng and Kumar, Sonal and Sakshi, S and Kim, Jaehyeon and Ping, Wei and Valle, Rafael and Manocha, Dinesh and Catanzaro, Bryan},
  journal={Proc. ICML},
  year={2025}
}

@inproceedings{elizalde2023clap,
  title={{CLAP} learning audio concepts from natural language supervision},
  author={Elizalde, Benjamin and Deshmukh, Soham and Al Ismail, Mahmoud and Wang, Huaming},
  booktitle={Proc. ICASSP},
  year={2023},
}

@article{li2023mert,
  title={{MERT}: Acoustic music understanding model with large-scale self-supervised training},
  author={Li, Yizhi and Yuan, Ruibin and Zhang, Ge and Ma, Yinghao and Chen, Xingran and Yin, Hanzhi and Xiao, Chenghao and Lin, Chenghua and Ragni, Anton and Benetos, Emmanouil and others},
  journal={Proc. ICLR},
  year={2024}
}

@article{deshmukh2025mellow,
  title={Mellow: a small audio language model for reasoning},
  author={Deshmukh, Soham and Dixit, Satvik and Singh, Rita and Raj, Bhiksha},
  journal={arXiv preprint arXiv:2503.08540},
  year={2025}
}

@article{gardner2023llark,
  title={{LLark}: A multimodal instruction-following language model for music},
  author={Gardner, Josh and Durand, Simon and Stoller, Daniel and Bittner, Rachel M},
  journal={Proc. ICML},
  year={2024}
}

@article{agostinelli2023musiclm,
  title={Music{LM}: Generating music from text},
  author={Agostinelli, Andrea and Denk, Timo I and Borsos, Zal{\'a}n and Engel, Jesse and Verzetti, Mauro and Caillon, Antoine and Huang, Qingqing and Jansen, Aren and Roberts, Adam and Tagliasacchi, Marco and others},
  journal={Proc. ICML},
  year={2023}
}

@inproceedings{law2009evaluation,
  title={Evaluation of algorithms using games: The case of music tagging.},
  author={Law, Edith and West, Kris and Mandel, Michael I and Bay, Mert and Downie, J Stephen},
  booktitle={ISMIR},
  year={2009},
}

@inproceedings{gemmeke2017audio,
  title={{AudioSet}: An ontology and human-labeled dataset for audio events},
  author={Gemmeke, Jort F and Ellis, Daniel PW and Freedman, Dylan and Jansen, Aren and Lawrence, Wade and Moore, R Channing and Plakal, Manoj and Ritter, Marvin},
  booktitle={Proc. ICASSP},
  year={2017},
}

@article{doh2023lp,
  title={{LP-MusicCaps}: {LLM}-based pseudo music captioning},
  author={Doh, SeungHeon and Choi, Keunwoo and Lee, Jongpil and Nam, Juhan},
  journal={arXiv preprint arXiv:2307.16372},
  year={2023}
}

@article{bai2024audiosetcaps,
  title={Audiosetcaps: An enriched audio-caption dataset using automated generation pipeline with large audio and language models},
  author={Bai, Jisheng and Liu, Haohe and Wang, Mou and Shi, Dongyuan and Wang, Wenwu and Plumbley, Mark D and Gan, Woon-Seng and Chen, Jianfeng},
  journal={IEEE Transactions on Audio, Speech and Language Processing},
  year={2025},
}

@inproceedings{banerjee2005meteor,
  title={{METEOR}: An automatic metric for MT evaluation with improved correlation with human judgments},
  author={Banerjee, Satanjeev and Lavie, Alon},
  booktitle={Proceedings of the acl workshop on intrinsic and extrinsic evaluation measures for machine translation and/or summarization},
  year={2005}
}

@article{zhang2019bertscore,
  title={{BertScore}: Evaluating text generation with bert},
  author={Zhang, Tianyi and Kishore, Varsha and Wu, Felix and Weinberger, Kilian Q and Artzi, Yoav},
  journal={arXiv preprint arXiv:1904.09675},
  year={2019}
}

@article{hu2022lora,
  title={Lora: Low-rank adaptation of large language models.},
  author={Hu, Edward J and Shen, Yelong and Wallis, Phillip and Allen-Zhu, Zeyuan and Li, Yuanzhi and Wang, Shean and Wang, Lu and Chen, Weizhu and others},
  journal={Proc. ICLR},
  year={2022}
}

@article{li2025reinforcement,
  title={Reinforcement learning outperforms supervised fine-tuning: A case study on audio question answering},
  author={Li, Gang and Liu, Jizhong and Dinkel, Heinrich and Niu, Yadong and Zhang, Junbo and Luan, Jian},
  journal={arXiv preprint arXiv:2503.11197},
  year={2025}
}

@article{dinkel2025midashenglm,
  title={{MiDashengLM}: Efficient Audio Understanding with General Audio Captions},
  author={Dinkel, Heinrich and Li, Gang and Liu, Jizhong and Luan, Jian and Niu, Yadong and Sun, Xingwei and Wang, Tianzi and Xiao, Qiyang and Zhang, Junbo and Zhou, Jiahao},
  journal={arXiv preprint arXiv:2508.03983},
  year={2025}
}

@article{quelennec2025matpac++,
  title={{MATPAC}++: Enhanced Masked Latent Prediction for Self-Supervised Audio Representation Learning},
  author={Quelennec, Aurian and Chouteau, Pierre and Peeters, Geoffroy and Essid, Slim},
  journal={arXiv preprint arXiv:2508.12709},
  year={2025}
}

@article{defferrard2016fma,
  title={{FMA}: A dataset for music analysis},
  author={Defferrard, Micha{\"e}l and Benzi, Kirell and Vandergheynst, Pierre and Bresson, Xavier},
  journal={Proc. ISMIR},
  year={2016}
}

@article{chen2022beats,
  title={Beats: Audio pre-training with acoustic tokenizers},
  author={Chen, Sanyuan and Wu, Yu and Wang, Chengyi and Liu, Shujie and Tompkins, Daniel and Chen, Zhuo and Wei, Furu},
  journal={arXiv preprint arXiv:2212.09058},
  year={2022}
}

@inproceedings{george2001automatic,
  title={Automatic musical genre classification of audio signals},
  author={George, Tzanetakis and Georg, Essl and Perry, Cook},
  booktitle={Proc. ISMIR},
  year={2001}
}

@misc{lostanlen2019medley,
  author       = {Lostanlen, Vincent and Cella, Carlos-Emilio and Bittner, Rachel and Essid, Slim},
  title        = {Medley-solos-DB: A cross-collection dataset for musical instrument recognition},
  year         = {2019},
  howpublished = {\url{https://doi.org/10.5281/zenodo.1344103}},
  doi          = {10.5281/zenodo.1344103}
}

\end{document}